\documentclass[reprint,secnumarabic,amssymb,amsmath,aps,prb,groupedaddress,frontmatterverbose,showpacs]{revtex4-1}
\usepackage[dvips]{graphicx}
\usepackage{bm}
\usepackage{color}
\begin{document}
\title{Observation of non-Fermi liquid behavior in hole doped Eu$_2$Ir$_2$O$_7$ }
\author{A. Banerjee$^1$, J. Sannigrahi$^2$, S. Giri$^1$, S. Majumdar$^1$}
\email{sspsm2@iacs.res.in}
\affiliation{$^1$Department of Solid State Physics, Indian Association for the Cultivation of Science, 2A \& B Raja S. C. Mullick Road, Jadavpur, Kolkata 700 032, India}
\affiliation{$^2$ISIS Neutron and Muon Source Science and Technology Facilities Council, Rutherford Appleton Laboratory, Didcot
OX11 0QX ,United Kingdom}

\begin{abstract}
The Weyl semimetallic compound Eu$_2$Ir$_2$O$_7$ along with its hole doped derivatives (which is achieved by substituting trivalent Eu by divalent Sr) are investigated through  transport, magnetic and calorimetric  studies. The metal-insulator transition (MIT) temperature is found to get substantially reduced with hole doping and for 10\% Sr doping the composition is metallic down to temperature as low as 5 K. These doped compounds are found to violate the Mott-Ioffe-Regel condition for minimum electrical conductivity and show distinct signature of non-Fermi liquid behavior at low temperature. The MIT in the doped compounds does not correlate with the magnetic transition point and Anderson-Mott type disorder induced localization may be attributed to the ground state insulating phase. The observed non-Fermi liquid behavior can be understood on the basis of disorder induced distribution of spin orbit coupling parameter which is markedly different in case of Ir$^{4+}$ and Ir$^{5+}$ ions.    

\end{abstract} 
\maketitle

\par
Recently the physics of Iridium based oxides have created considerable excitement due to their fascinating electronic and magnetic properties originating from the interplay between  spin-orbit coupling (SOC), electron correlation ($U$) and inter-site charge hopping.~\cite{rotenberg, arima} On account of the extended nature of the 5$d$ orbitals of Ir, these materials are associated with wider electronic band ($W$) and relatively weaker $U$ and are expected to be metallic in nature. However due to large atomic number of Ir, the relativistic SOC plays an important role (in fact the energy scales of SOC and $U$ are comparable in magnitude) which in tandem with $U$ can lead to an insulating ground state. The classic example is the layered oxide Sr$_2$IrO$_4$ with Ir$^{4+}$ (5d$^{5}$),~\cite{sr2iro4} where SOC splits the $t_{2g}$ level into two bands with  pseudo spins $J_{eff}$ = $\frac{1}{2}$ doublet and a $J_{eff}$ = $\frac{3}{2}$ quadruplet and thereby effectively reduces the bandwidth [see fig. 1(a)]. The higher occupied  $J_{eff}$ = $\frac{1}{2}$ band is half filled and can open up a Mott gap through electron correlation. However, the system can have a metallic ground state if the band width of the $t_{2g}$ level is large [see fig. 1 (b)].

\par
Apart from Sr$_2$IrO$_4$, the pyrochlore iridates $R_2$Ir$_2$O$_7$ (R = rare-earth) are found to be quite intriguing material where the samples show metal-insulator (MI) transition on cooling leading to an insulating ground state.~\cite{hinatsu, nakatsuji} Such MI transition is expected to be associated with the interplay between SOC and electron correlations. For decades, rare-earth-pyrochlores such as Dy$_2$Ti$_2$O$_7$ or Ho$_2$Ti$_2$O$_7$  have attracted  considerable attention for their unique magnetic frustration of geometrical origin  which leads to spin-ice state and the formation of Dirac monopoles.~\cite{spin-ice, monopole} The pyrochlore iridates are also correlated electron system showing magnetic frustration. Theoretical studies indicate that these compounds are good candidates for topological materials.~\cite{pesin,yang,wan} Very recently Sushkov {\it et al.} identified Eu$_2$Ir$_2$O$_7$ to be a Weyl Semimetal from the THz optical conductivity studies.~\cite{dassarma}

\begin{figure}[t]
\centering
\includegraphics[width = 9 cm]{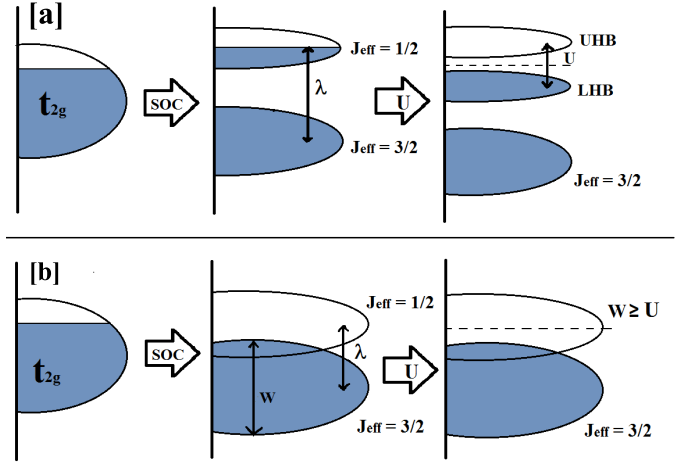}
\caption {(Color online) (a) provides a schematic description of the splitting of $t_{2g}$ band of Ir-5$d$ under  SOC and $U$ leading to an insulating state where  completely filled  lower Hubbard band (LHB) is separated from the empty upper Hubbard band (UHB) by an energy gap. (b) shows the case with large bandwidth ($W \gtrsim$ SOC, $U$), where overlapping bands can lead to a metallic state (see text).}
\end{figure}

\par
The MI transition temperature ($T_{MI}$) in $R_2$Ir$_2$O$_7$ varies monotonically with the ionic radius of the $R$ atom.~\cite{hinatsu,matsu} $T_{MI}$ also coincides with the magnetic transition temperature ($T_N$) of the compounds and it is found to be second order in nature.~\cite{matsu} At $T_N$,  they undergo a commensurate long range order where Ir$^{4+}$ moments form a {\it all-in-all-out} configuration with all the  moments  pointing either toward or outward from the center of each tetrahedron.~\cite{tomiyasu,disseler,sagyama} It is to be noted that for $R_2$Ir$_2$O$_7$, both $R$ and Ir form corner sharing tetrahedra. Pesin and Balents~\cite{pesin} pointed out that the insulating ground state of these pyrochlores can be aptly described as `topological Mott' insulating phase. Ueda {\it et al.}~\cite{ueda} observed  successive phase change of the ground state of Nd$_2$Ir$_2$O$_7$ on doping from a narrow-gap Mott insulator to  Weyl semimetal, and  finally to a correlated metal, which can be attributed to the mutual interplay of SOC and $U$. 
 
\par
Despite the wealth of theoretical and experimental works, the true nature of the MI transition and the insulating ground state of these pyrochlores is a matter of debate. It is clear that the electron occupancy in Ir plays an important role towards the MI transition. In the present work we have performed a systematic study of hole doping in one such Ir-pyrochlore Eu$_2$Ir$_2$O$_7$ and performed transport, magnetic and calorimetric studies on the derived materials. Hole doping is expected to vary the Ir valency as well as it may affect the Ir-O-Ir bond angle. Our study indicates a systematic lowering  in $T_{MI}$ of the samples with Sr substitution. A significant outcome of the work is the observation of non-Fermi liquid ground state in certain samples, a phenomenon that was earlier predicted theoretically for topological states with strong SOC.~\cite{moon}

\begin{figure}[t]
\centering
\includegraphics[width = 9 cm]{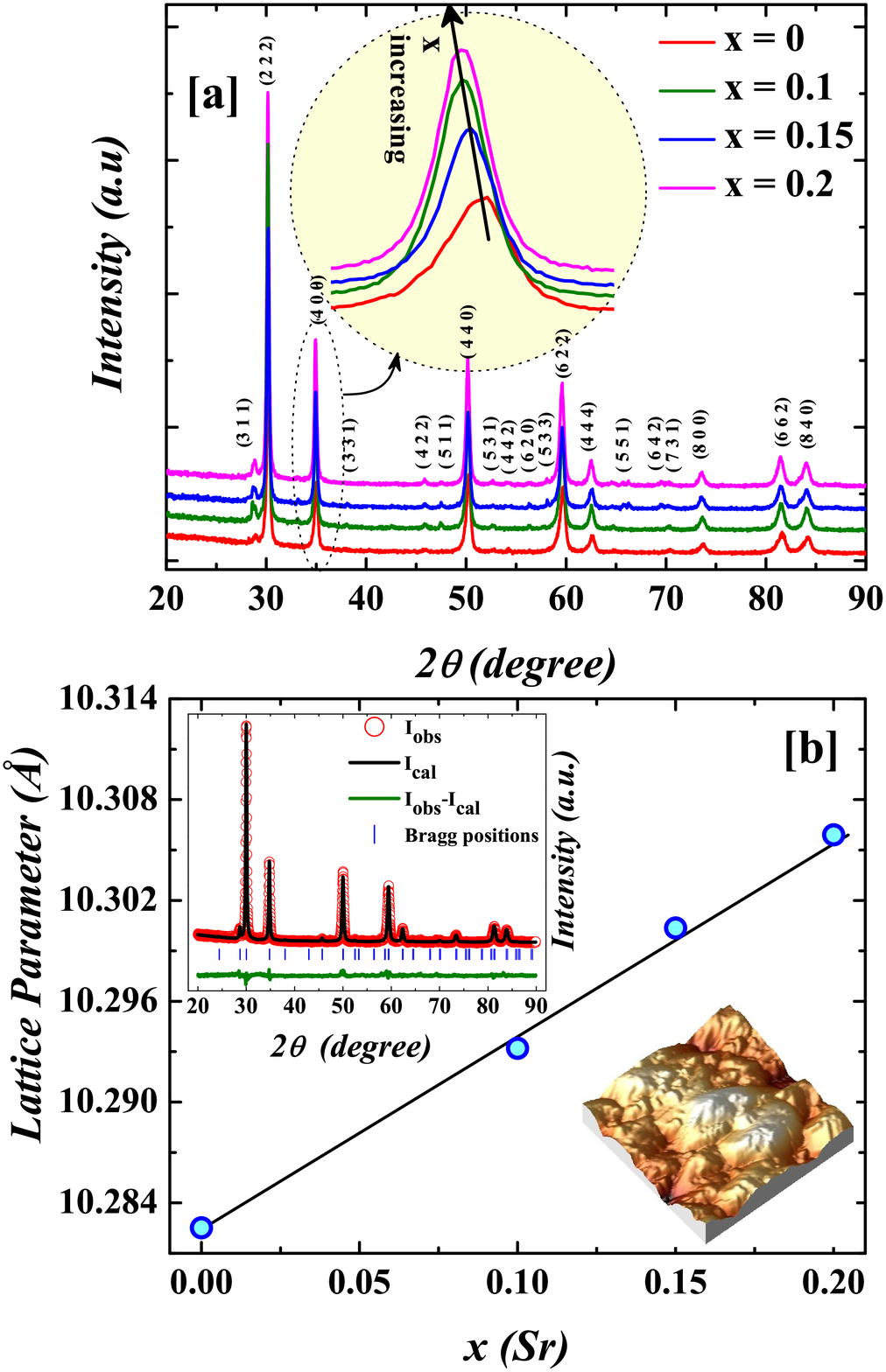}
\caption{(Color online) (a) shows the powder x-ray diffraction pattern of Eu$_{1-x}$Sr$_x$Ir$_2$O$_7$ ($x$=0.0, 0.1, 0.15 and 0.2) samples recorded at room temperature. The inset shows the shift of the (400) reflection with doping. (b) shows the variation of cubic lattice parameter as a function of doping concentration. The upper inset shows the Rietveld refined data for $x$ = 0.2 sample, where the solid line is the fitted curve and the scattered points are the experimental data. The lower inset of (b) shows the topograph of a flat surface of $x$ = 0.0 sample recorded by using atomic force microscope.}
\end{figure}

\par
Hole doping  in Eu$_2$Ir$_2$O$_7$ was realized by substituting Eu$^{3+}$ by Sr$^{2+}$. The polycrystalline samples Eu$_{1-x}$Sr$_x$Ir$_2$O$_7$ ($x$=0.0, 0.1, 0.15 and 0.2) were prepared through solid state reaction route. Stoichiometric amounts of Eu$_2$O$_3$, SrCO$_3$ and IrO$_2$ were mixed intimately and fired at 1223 K for one day. After pressing the mixture powder into pellets, they  were heated in air for 3 days at 1273 K with several intermediate grindings. The structure and phase purity of the samples were investigated by powder x-ray diffraction (XRD) using Cu K$_{\alpha}$ radiation. Detailed analysis confirms the cubic pyrochlore-type phase (space group $Fd\bar{3}m$) for all the samples as depicted in fig. 2 (a). The peaks shift to lower angle with increasing $x$, which is better viewed in the inset of fig. 2(a). We have performed Rietveld refinement of the XRD data using Fullprof suit~\cite{fp} and a representative fitted data is shown for $x$ = 0.2 sample in the upper inset of fig. 2 (b). The fitting for all the samples converges well with $\chi^2$ lying between $\sim$ 2 to 1. The site occupacy of different atoms are found to be close to unity indicating the absence of any  substantial site vacancies.  The  cubic lattice parameter ($a_c$) shows almost linear variation with $x$ following Vegard’s law [Fig. 2(b)], which indicates that Sr is systematically getting substituted at the Eu site.  In order to probe the grain and grain boundaries as well as any void in the sample, we studied the sample topography using atomic force microscope (Veeco-diCP II). A representative figure of a flat surface of  $x$ = 0.0 pellet is shown in the lower inset of fig. 2 (b). From the image, the grains are clearly visible measuring about 100-300 nm. The grains are closely packed without any obvious void. Similar topography is also obtained for the other samples.  The estimated mass density of the pellet of $x$ = 0.0 sample is about 7.3 ($\pm$ 10) g/cm$^3$, which about 75\% of the x-ray density and it is an indication of the compactness of the sample.  The magnetic measurements were performed on a vibrating sample magnetometer from Cryogenic Ltd. UK, as well as on a Quantum Design SQUID magnetometer. The zero-field and in-field resistivity ($\rho$) were measured by four-probe method on a cryogen-free high magnetic field system (Cryogenic Ltd., U.K.) with magnetic field ($H$) as high as 50 kOe and within the temperature ($T$) range between 2 and 300K.  Heat capacity ($C_p$) measurement was carried out using a Quantum Design physical properties measurement system. The pure and $x$ = 0.2 samples were also investigated through core level x-ray photo-electron spectroscopy (XPS) at room temperature using Al K$_\alpha$ radiation on a laboratory based commercial instrument (Omicron).  The sample surfaces were cleaned {\it in situ}~by argon ion  sputtering. Fig. 3 (a) and (b) show the spin-orbit split Ir-4$f$ core level spectra for $x$ = 0.0 and 0.2 samples respectively. The spectrum for  $x$ = 0.2  sample shifts to higher binding energy presumably due to the presence of Ir$^{5+}$ states. The spectrum of the pure sample can be fitted with a single spin-orbit split doublet with a separation of 3.0 eV, while for the $x$ = 0.2 sample two doublets are required to achieve good fitting. The additional weak doublet (marked in horizontal lines) occurring at a higher binding energy is likely to be connected with the minority Ir$^{5+}$ ions,~\cite{Ir5} indicating the mixed valency in the Sr doped sample.       

\begin{figure}[t]
\centering
\includegraphics[width = 9 cm]{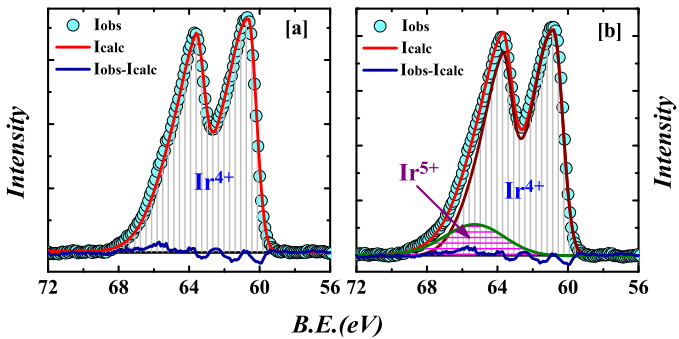}
\caption{(Color online) (a) and (b) show the 4$f$ core level XPS data for $x$ = 0.0 and $x$ = 0.2 samples respectively. The solid lines are fit to the data with the model discussed in the text.}

\end{figure}

\par
Figs. 4 (a-d)  show the $T$ dependence of magnetic susceptibility $\chi (T)$ ($= M(T)/H$) for $x$=0.0, 0.1, 0.15 and 0.2 samples measured in zero-field-cooled (ZFC) and field-cooled (FC) protocols. For the parent sample, a magnetic anomaly is observed close to  120 K, below which there is a clear bifurcation between the ZFC and FC $\chi$ and it matches well with literature.~\cite{matsu,nakatsuji} It is known  that  several Ir-pyrochlores show long range  {\it all-in-all-out} type of magnetic order of Ir close to such thermo-magnetic irreversibility ({\it i.e.,} the bifurcation point of FC and ZFC $\chi$).~\cite{matsu,tomiyasu} The point of thermo-magnetic irreversibility is often found to be different from actual ordering temperature, however it provides a rough indication of the onset of an ordered magnetic phase.  We find that the  bifurcation is also present in the Sr substituted samples [figs. 4 (b-d)], and the extend of bifurcation ({\it i.e.} $M_{FC}-M_{ZFC}$) diminishes with increasing Sr concentration. The bifurcation point also shifts to lower temperature [indicated by the arrows in fig. 4 (a-d)], and they are found to be $T_{IR}$ $\sim$ 120, 107, 97 and 60 K respectively for $x$ = 0.0, 0.1, 0.15 and 0.2 samples. 
\par
The $T$ variation of heat capacity ($C_p$) shows a peak at 99 K [fig. 4 (e)] for $x$ = 0.2 sample, which can be ascribed to be the $T_N$ of the sample. This value of $T_N$ obtained from heat capacity is higher than $T_{IR}$ of the sample. Notably, the feature at $T_N$ is distinct but weak. However, this weak nature is  very similar to the data obtained for the single crystalline sample of Eu$_2$Ir$_2$O$_7$.~\cite{nakatsuji}. In order to rule out the possible effect of impurity responsible for the weak feature in the $C_p(T)$ data, we calculated the magnetic entropy ($S_m$) as depicted in fig. 4(e) (inset) as a function of $T$. In absence of suitable nonmagnetic counterpart, we have fitted the high-$T$ data (above $T_N$) with the Debye heat capacity model and an electronic part ($C_{nm}(T) = C_{Debye}(\Theta, T) + \gamma T$, where $\Theta$ is the Debye temperature and $\gamma$ is the Sommerfeld coefficient), and obtained the magnetic part, $C_m = C_p - C_{nm}$. Subsequently, $S_{m}$ was calculated by using the formula $S_m(T) = \int_0^T(C_m/T^{\prime})dT^{\prime}$. We see that $S_m(T)$ tends to saturate above about 100 K, with a value of 34 J/mol-K. The expected value of $S_m$ for the low spin state ($J_{eff} = \frac{1}{2}$) is much lower ($\sim$ 11 J/mol-K). Such discrepancy may due to a poor estimation of lattice heat capacity or from an contribution from the induced moment at the Eu site. In fact our analysis of the $\chi(T)$ data above 175 K (in the paramagnetic state), indicates a large effective paramagnetic moment ($p_{eff}$ = 7.652 $\mu_B$/formula unit), which cannot be accounted by the low spin $J_{eff} = \frac{1}{2}$ state.

\par
Thermo-magnetic irreversibility in a material can depend upon various factors such as disorder, magnetic anisotropy etc. In the present case, $T_{IR}$ albeit lies below $T_N$, it signifies the presence of a magnetic state at least below the bifurcation point.  The isothermal magnetization curves ($M$ vs $H$) for all the samples measured at 4 K show  linear behavior without any coercivity.  As an example, the $M-H$ curve for $x$ = 0.0 sample for $H = \pm$ 150 kOe is depicted in fig. 4(g), which is found to be perfectly linear even at higher fields. This does not support the spin canted ground state predicted for  Eu$_2$Ir$_2$O$_7$ compounds~\cite{nakatsuji} and rather indicates an antiferromagnetic (AFM) ground state for parent and doped compositions. Such conclusion is based on the fact that had it been a spin canted system, there would  be a nonvanishing FM component. This would certainly be reflected in the $M-H$ curve in the form of hysteresis and/or a tendency of saturation of $M$ at higher values of $H$.  On the other hand, a rise in FC $\chi$ is observed for all the samples below $T_{IR}$, which is not generally expected for an AFM system. A very recent experimental work based on torque magnetometry on Eu$_2$Ir$_2$O$_7$ indicates the  development of a perpendicular magnetization ($M_{\bot}$) of octupolar origin on application of magnetic field in the $a-b$ plane of the crystal.~\cite{liang} The rise in $\chi$ below $T_{IR}$ may be related to this $M_{\bot}$.

\par
The MI transition in all the samples is probed by dc resistivity studies. The parent sample shows MI transition at $T_{MI}$ = 120 K below which the sample is insulating. Most interestingly, we found a huge but systematic decrease  in $T_{MI}$  with the increase of doping percentage. T$_{MI}$ is found to be  120 K, 78 K, 38 K and 5 K for $x$ = 0.0, 0.1, 0.15 and 0.2 samples respectively [figs. 5(a-d)]. The variation of $T_{MI}$ is found to be almost linear with $x$ [fig. 6 (a)]. Our observation points to some important facts regarding $T_{MI}$. Firstly, $T_{MI}$ of the samples does not directly corresponds to the magnetic transition temperature. For example, $x$ = 0.2 sample has T$_N$ close to 99 K, while $T_{MI}$ is far below than this. This rules out a simple Mott-Slater kind of insulating ground state of the sample. It is to be noted that $T_{MI}$ in $R_2$Ir$_2$O$_7$ for different $R$ atoms corresponds exactly to their respective $T_N$.~\cite{matsu} Secondly, $T_{MI}$ of these samples is found to be insensitive to the applied magnetic field [see inset of fig. 5 (a) and (b), for $x$ = 0.0 and 0.1 samples respectively], which is in contrast with the the scenario of 3$d$ system such as perovskite manganites where $T_{MI}$ is strongly field dependent. We have also plotted the variation of $T_{IR}$ with $x$ [fig. 6(b)] and it is clearly seen that  $T_{IR}$'s for different samples lie well above $T_{MI}$.

\par
 In order to rule out the effect of oxygen off-stoichiometry, we annealed a sample piece ($x$ = 0.0) in oxygen flow at 800$^{\circ}$ C and measured $\rho$ as a function of $T$. We do not find any noticeable difference in the $\rho(T)$ behavior before and after oxygen annealing (plots not shown hre). It turns out that oxygen off-stoichiometry, even if present to some extend,  is not playing any role in altering $T_{MI}$.  It is to be noted that for Nd$_2$Ir$_2$O$_7$ thin films, oxygen annealing has a marked effect on the physical properties.~\cite{nd2ir2o7-film}

\begin{figure}[t]
\centering
\includegraphics[width = 9 cm]{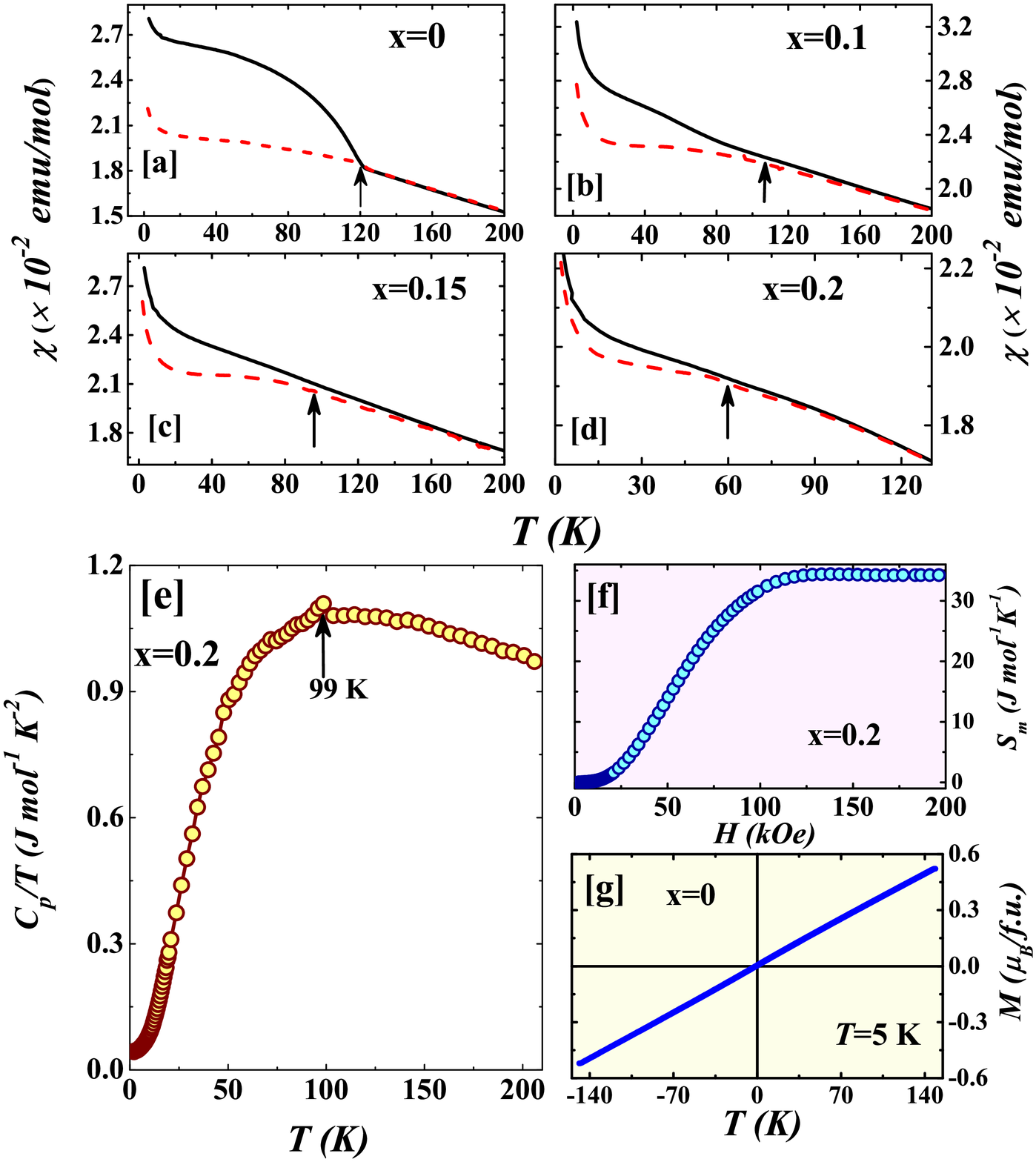}
\caption {(Color online) (a-d) show DC magnetic susceptibility ($\chi$) as a function of temperature for Eu$_{1-x}$Sr$_x$Ir$_2$O$_7$ ($x$=0.0, 0.1, 0.15 and 0.2)  at 2 kOe in ZFC and FC protocols. (e) represents the heat capacity as a function of temperature for $x$ = 0.2 sample. The thermal variation of magnetic entropy is shown in (f), while (g) represents the isothermal magnetization data recorded at 5 K up to the applied  field as high as 150 kOe.}
\end{figure}

\begin{figure}[t]
\centering
\includegraphics[width = 9 cm]{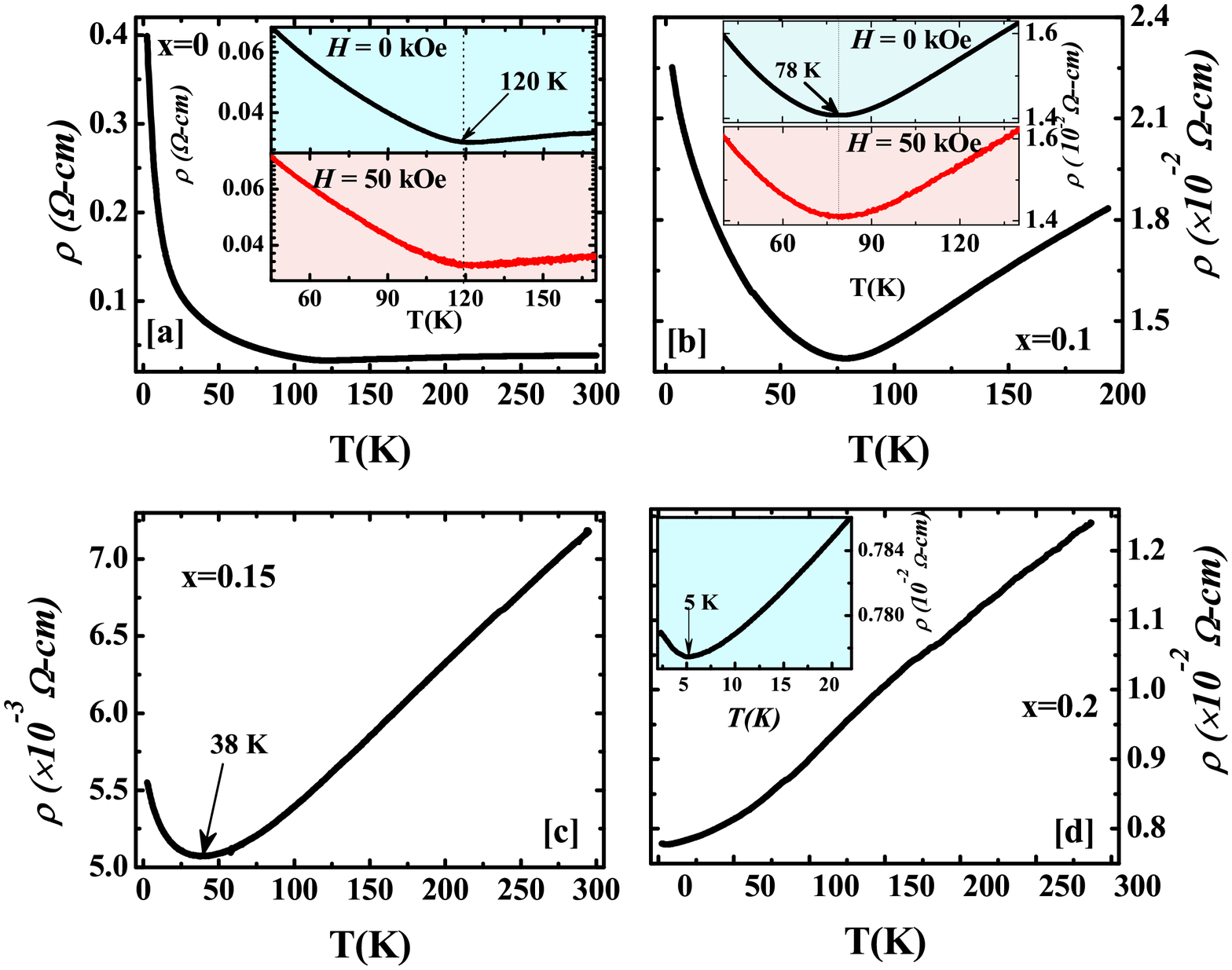}
\caption {(Color online) (a-d) present the resistivity data as a function of temperature for Eu$_{1-x}$Sr$_x$Ir$_2$O$_7$. The insets of (a) and (b)  show the metal-insulator transition point at zero and 50 kOe of applied magnetic fields for $x$ = 0.0 and 0.1 samples respectively. The inset in (d) shows an enlarged view of the metal-insulator transition for $x$ = 0.2.}
\end{figure}

\par
Considering the fact that $x$ = 0.2 sample shows MI transition at a fairly low $T$, we looked carefully the $T$ dependence of $\rho$ between 7 and 44 K. It is evident that despite the system has a metallic character, $\rho$ does not show a clear $T^2$ variation [fig. 7 (a)] as expected for a Fermi liquid. The $\rho(T)$ data can be better described by an empirical $T^{1.5}$ variation as depicted in fig. 7 (b). Clearly, there is a downward curvature at the low-$T$ side in fig. 7 (a), which is absent in fig. 7 (b). We have also looked at the low temperature heat capacity data of the samples and fig.7 (c) and (d) respectively show the $C_p/T$ versus $T^2$ data below 20 K for $x$ = 0.0 and 0.2 samples. We do not see any signature of $T_{MI}$ in the $C_p$ data for the $x$ = 0.2 sample at 5 K. The parent sample shows a linear $C_p/T$ versus $T^2$ plot which is expected for a crystalline solid as total heat capacity $C_p$ = $\gamma T$ + $\beta$ $T^3$, where $\gamma T$  and $\beta$ $T^3$ are respectively electronic and lattice parts. Being an AFM insulator, the curve passes through the origin, indicating that there is negligible electronic contribution to the heat capacity. The spin wave contribution towards $C_p$  for an AFM system also varies as $T^3$ and it is indistinguishable from the lattice part. On the other hand for $x$ = 0.2, the $C_p/T$ versus $T^2$ data deviates strongly from the linear behavior below about 12 K. Such observation is at par with the non-quadratic $T$-dependance of the $\rho$ data at low temperature. The magnetic contribution to $C_p$ is likely to be both AFM in $x$ = 0 and 0.2 samples (as evident from our magnetic measurements), and the deviation from linearity in the $C_p/T$ versus $T^2$ data is unlikely to be associated with the spin wave contribution. Also, the spin wave part is generally negligible below 10 K ($T_N$ is of the order of 100 K). Such observation clearly demonstrates that the material shows unconventional metallic character. The low $T$ data can be fitted empirically quite well  with a relation $C_p = \gamma^{\prime} T \ln{(T_0/T)} + \beta T^3$ [see inset of fig. 7 (d)], which indicates a logarithmic variation of the density of states at low temperature. In our fitting we find that $T_0$ =17.9 K, which is much higher than the $T_{MI}$ of the sample. 

\begin{figure}[t]
\centering
\includegraphics[width = 9 cm]{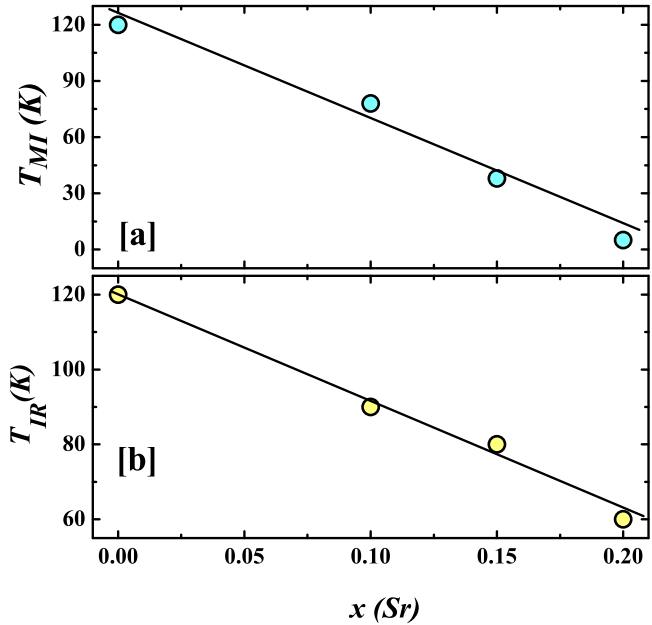}
\caption {(Color online) (a) and (b) respectively  show the variation of $T_{MI}$ and $T_{IR}$ with $x$}
\end{figure}

\begin{figure}[t]
\centering
\includegraphics[width = 9 cm]{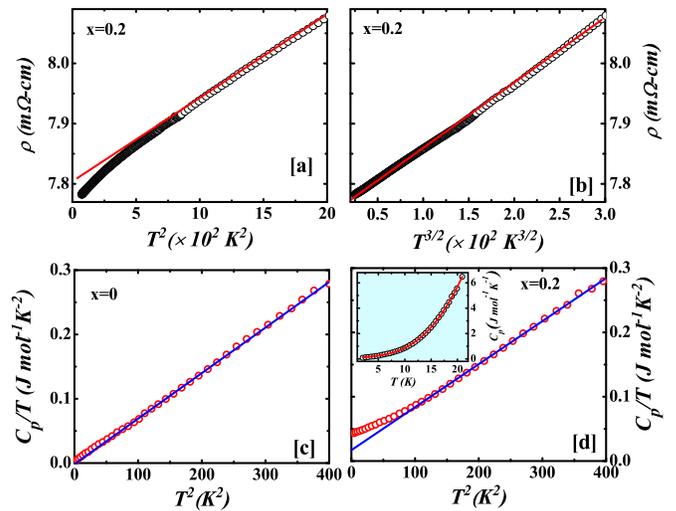}
\caption {(Color online) (a) and (b) show respectively the variation of low temperature resistivity as a function of $T{^2}$ and $T^{3/2}$ for $x$ = 0.2. The solid  lines are for guide to the eye. (c) and (d) show the variation of $C_p/T$ with $T^2$ below 20 K for $x$ = 0.0 and 0.2 samples respectively. The inset shows the low temperature $C_p$ versus $T$ data along with a fit (solid line) with the relation $C_p = \gamma^{\prime} T \ln{(T_0/T)} + \beta T^3$ }
\end{figure}

\par
The plausible scenario for the metallicity in the Sr substituted materials is connected to hole doping which turns the sample mixed valent with the presence of both Ir$^{4+}$ and Ir$^{5+}$ ions. It is well established that SOC is lower in case of Ir$^{5+}$ (5$d^{4}$) as compared to Ir$^{4+}$ (5$d^{5}$)~\cite{khomskii}. In a recent work by Bhowal et al.~\cite{bhowal} based on density functional based calculations, it is found that for double perovskite based $d^{4}$-iridates, the electronic bandwidth, $W$ of the $t_{2g}$ level is higher. In the event of  $W \gtrsim$ SOC, there can be overlap of $J_{eff}$ = 1/2 and $J_{eff}$ = 3/2 bands. If $W \gtrsim$ (SOC and $U$), there can exist partially filled Hubbard bands even in presence of $U$. Further, hybridization between $d^{5}$ and $d^{4}$ states can lead to additional band widening in the doped samples [fig. 1 (b)] . Such effects can together facilitate band overlapping leading to a more metallic behavior in the Sr doped samples. 
\par
The transitions at $T_{MI}$ for the doped (as well as the parent) compositions are found to be continuous (unlike pure Mott transition), and do not correspond grossly to $T_N$. Therefore, Anderson localization type phenomenon may be the prevalent cause for the insulating ground state even for the highest doped sample. Considering the presence of finite electronic correlation, the transition may aptly be described as an Anderson-Mott type.~\cite{belitz}

\par
One of the important observations is the non-Fermi liquid behavior of the highest doped sample ($x$ = 0.2) as evident from both $\rho$ and $C_p$ studies. The sample is already in the magnetically ordered state and proximity to a quantum critical point can be ruled out for NFL behavior.  In addition, it is unlikely for quantum fluctuation to dominate over thermal fluctuation at 10 K (the temperature below which NFL state is observed).   Despite the metallic nature down to 5 K in $x$ = 0.2, the absolute value of $\rho$ is high. The Mott-Ioffe-Regel (MIR) value of minimum conductivity of a metal is given by $\sigma_{min} =$  0.03$e^2/\hbar r_{nn}$,~\cite{mott}  where $r_{nn}$ is the nearest neighbor atomic distance and for pyrochlore structure it is ($a_c$/4)$\sqrt{2}$.  For $x$ = 0.2 sample, the maximum metallic resistivity $\rho_{max}$ (reciprocal of $\sigma_{min}$)  turns out to be 4.8 m$\Omega$-cm, which is lower than the experimentally observed $\rho$ even at 5 K with out the signature of {\it resistivity saturation}. Such {\it bad metals} violating the MIR rule are not uncommon among  correlated electron systems including high $T_C$ cuprates, ruthenates, nickelates and doped fullerides.~\cite{emery,cao,ir,mikheev} and a mutual interplay between disorder and correlation has often been attributed to the effect. Notably, the MIR formula had been  primarily derived for disordered solids (so called bad metals) and is equally applicable for polycrystalline sample with grain and grain boundaries.~\cite{edwards}

\par
The deviation from an FL ground state and the violation of MIR rule can be thought to be inter-correlated,~\cite{emery} however it is not universally supported by the experimental observations.~\cite{philmag} There are examples of bad metals among ruthenates which show FL characteristics.~\cite{cao,andy} The  metallic NFL behavior is conjectured  theoretically in a correlated disordered system prior to an Anderson-Mott type MI transition.~\cite{kotliar} The situation might be more exciting in these doped Eu$_2$Ir$_2$O$_7$ compositions where SOC is also important in addition to $U$. The presence of NFL state in correlated electron system with SOC (such as pyrochlore iridate) has been argued by Moon et al.~\cite{moon} In our study, the  NFL behavior is only observed in the doped composition and this may be an indication of the role of disorder towards the observed deviation from FL. In fact, NFL behavior in the metallic phase of  SrIrO$_3$ thin films is attributed to the strain induced disorder.~\cite{biswas} Effect of disorder is well documented in Kondo lattice systems, where a distribution of Kondo temperature can lead to an NFL ground state.~\cite{kondo} Since, SOC in Ir$^{4+}$ is higher than Ir$^{5+}$ ion~\cite{khomskii}, one can have a spatial distribution of SOC over the doped sample, which can be held responsible for the NFL behavior.

\par
Summarizing, we find a systematic change in the metal-insulator transition temperature on hole doping in Eu$_2$Ir$_2$O$_7$. From our sample characterization by atomic force microscope, these polycrystalline ceramics samples are found to be made up of closely packed grains without much void spaces. The careful analysis of the x-ray diffraction pattern also indicate full site occupancy of the atoms at the crystallographic positions. The oxygen off-stoichiometry can also be ruled out for the change in $T_{MI}$, as additional oxygen annealing has little or no effect on the metal-insulator transition. The x-ray photoemission spectra indicate clear presence of Ir$^{5+}$, which is playing an important role in determining ground state electronic and magnetic properties. We find that the system violets the Mott-Ioffe-Regel value of minimum conductivity for disordered bad metals, which indicates that the strong electronic correlation in presence of disorder is operating here. The most important observation is the NFL behavior in case of $x$ = 0.2 sample. Such deviation from the regular FL ground state cannot be ascribed to polycrystalline nature of the sample (grain, grain boundaries or other dislocations), as it is absent in the parent compound. We opine that the random distribution of Ir$^{4+}$ and Ir$^{5+}$ ions in the sample provides a spatial variation of  spin-orbit coupling strength, which in turn provides multiple  renormalization coefficients for the interacting electrons  to be mapped on a simple non-interacting Drude-Sommerfeld model to form the FL state. Such multiplicity as a whole breaks the FL state, and an NFL state emerges out. The NFL state, thus indirectly depends upon the disorder at the atomic level, but not on the bulk disorder. It resembles the Kondo disorder model of some heavy fermion alloys showing NFL, where a distribution Kondo temperature (as compared to SOC energy scale in the present case) breaks the FL state.

\par
The authors would like to thank I. Dasgupta and S. Bhowal (IACS) for useful discussions. K. G. Suresh (IIT Bombay) and S. Chatterjee (UGC-DAE-CSR, Kolkata) are thankfully acknowledged for different measurements. AB wishes to thank DST-INSPIRE program for the research assistance. JS wishes to acknowledge  EU's Horizon 2020 research and innovation programme under the Marie Sk\l{odowska}-Curie grant agreement (No 665593) awarded to the STFC, UK.


\begin{thebibliography}{99}

\bibitem{rotenberg} B. J. Kim, H. Jin, S. J.Moon, J.-Y. Kim, B.-G. Park, C. S. Leem, J. Yu, T. W. Noh, C. Kim, S.-J. Oh, J.-H. Park, V. Durairaj, G.Cao and E. Rotenberg, Phys. Rev. Lett. {\bf 101} 076402 (2008).

\bibitem{arima} B. J. Kim, H. Ohsumi, T. Komesu, S. Sakai, T.Morita, H. Takagi and T. H. Arima,  Science {\bf 323} 1329 (2009).

\bibitem{sr2iro4} B. J. Kim, H. Jin, S. J. Moon, J.-Y. Kim, B.-G. Park, C. S. Leem, Jaejun Yu, T.W. Noh, C. Kim, S.-J. Oh,
J.-H. Park, V. Durairaj, G. Cao, and E. Rotenberg, Phys. Rev. Lett. {\bf 101}, 076402 (2008)


\bibitem{hinatsu} K. Matsuhira, M. Wakeshima, R. Nakanishi, T. Yamada, A. Nakamura,W. Kawano, S. Takagi and Y. Hinatsu, J. Phys. Soc.
Jpn. {\bf 76} 043706 (2007) 

\bibitem{nakatsuji} J. J. Ishikawa, E. C. T. O’Farrell and S. Nakatsuji, Phys. Rev. B {\bf 85} 245109 (2012).

\bibitem{spin-ice} S. T. Bramwell, and M. J. P. Gingras, Science {\bf 294}, 1495 (2001).

\bibitem{monopole} L. D. C. Jaubert,  and P. C.W. Holdsworth, Nat. Phys. {\bf 5}, 258 (2009); S. Ladak, D. E. Read, G. K. Perkins, L. F. Cohen and W. R. Branford, {\it ibid} {\bf 6}, 359 (2010). 

\bibitem{pesin} D. Pesin, and Leon Balents, Nat. Phys. {\bf 6}, 376 (2010).

\bibitem{yang} B.-J. Yang, and Yong Baek Kim, Phys. Rev. B {\bf 82}, 085111 (2010).

\bibitem{wan} X. Wan, A. M. Turner, A. Vishwanath,and S. Y. Savrasov, Phys. Rev. B {\bf 83}, 205101 (2011) 


\bibitem{dassarma} A. B. Sushkov, J. B. Hofmann, G. S. Jenkins, J. Ishikawa, S. Nakatsuji,  S. Das Sarma, and H. D. Drew,  Phys. Rev. B {\bf 92}, 241108(R) (2015).

\bibitem{matsu} K. Matsuhira, M. Wakeshima, Y. Hinatsu, and S. Takagi, J. Phys.
Soc. Jpn. {\bf 80}, 094701 (2011).

\bibitem{tomiyasu} K. Tomiyasu, K. Matsuhira, K. Iwasa, M. Watahiki, S. Takagi, M. Wakeshima, Y. Hinatsu, M. Yokoyama, K. Ohoyama, and
K. Yamada, J. Phys. Soc. Jpn. 81, 034709 (2012).

\bibitem{disseler} S. M. Disseler, C. Dhital, A. Amato, S. R. Giblin, C. de la Cruz, S. D. Wilson, and M. J. Graf, Phys. Rev. B {\bf 86}, 014428 (2012)

\bibitem {sagyama} H. Sagayama, D. Uematsu, T. Arima, K. Sugimoto, J. J. Ishikawa, E. O’Farrell, and S. Nakatsuji, Phys. Rev. B {\bf 87}, 100403(R) (2013).

\bibitem{ueda} K. Ueda, J. Fujioka, Y. Takahashi, T. Suzuki, S. Ishiwata, Y. Taguchi, and Y. Tokura, Phys. Rev. Lett. {\bf 109}, 136402 (2012).

\bibitem{moon} E.-G. Moon, C. Xu, Y. B. Kim, and L. Balents, Phys. Rev. Lett {\bf 111}, 206401 (2013).

\bibitem{fp} https://www.ill.eu/sites/fullprof/

\bibitem{Ir5} T. Otsubo, S. Takase, and Y. Shimizu, ECS Transactions {\bf 3}, 263 (2006).

\bibitem{liang} T. Liang, T. H. Hsieh, J. J. Ishikawa, S. Nakatsuji, L. Fu, and N. P. Ong, Nat. Phys. {\bf 13}, 599 (2017).

\bibitem{nd2ir2o7-film} J. C. Gallagher, B. D. Esser, R. Morrow, S. R. Dunsiger, R. E. A. Williams,
P. M. Woodward, D. W. McComb, and  F. Y. Yang, Sci. Rep. {\bf 6}, 22282 (2016)

\bibitem{khomskii} D. Khomskii, Transition Metal Compounds, (Cambridge University Press, 2014). 

\bibitem{bhowal} S. Bhowal, S. Baidya, I. Dasgupta, and T. Saha-Dasgupta, Phys. Rev. B {\bf 92}, 121113(R) (2015)


\bibitem{belitz} D. Belitz, and T. R. Kirkpatrick, Rev. Mod. Phys. 66, 261 (1994).

\bibitem{mott} N. F. Mott, {\it Metal-Insulator Transitions, 2/e} (Taylor \& Francis, London) (1974).


\bibitem{emery} V. J. Emery, and S.A. Kivelson, Phys. Rev. Lett. {\bf 74}, 3253 (1995).

\bibitem{cao} G. Cao, W.H. Song, Y.P. Sun, X.N. Lin, Solid State Commun. {\bf 131}, 331 (2004).

\bibitem{ir} O. Gunnarsson, M. Calandra, and J. E. Han, Rev. Mod. Phys. {\bf 75}, 1085 (2003).

\bibitem{mikheev} E. Mikheev, A. J. Hauser, B. Himmetoglu, N. E. Moreno, A. Janotti,
C. G. Van de Walle, S. Stemmer, Sci. Adv. {\bf 1}, e1500797 (2015).

\bibitem{edwards} P. P. Edwards, M. T. J. Lodge, F. Hensel, and R. Redmer, Phil. Trans. R. Soc. A  {\bf 368}, 941 (2010)


\bibitem{philmag} N. E. Hussey, K. Takenaka, and H. Takagi, Phil. Mag. {\bf 84}, 2847 (2004).

\bibitem{andy} A. P. Mackenzie, S. R. Julian, A. J. Diver, G. J. McMullan, M. P. Ray, G. G. Lonzarich, Y. Maeno,
S. Nishizaki, and T. Fujita, Phys. Rev. Lett. {\bf 76}, 3786 (1996).

\bibitem{kotliar} V. Dobrosavljevi\'c, and G. Kotliar,  Phys. Rev. Lett. {\bf 78}, 3943 (1997).

\bibitem{biswas} A. Biswas, K.-S. Kim, and Y. H. Jeong, J. Appl. Phys. {\bf 117}, 115304 (2015).

\bibitem{kondo} E. Miranda and V. Dobrosavljevi\'c, and G. Kotliar, Phys. Rev. Lett. {\bf 78}, 290 (1997). 


 

\end{thebibliography}
\end{document}